\documentclass[manuscript]{rmaa}
\usepackage{psfrag,color}
\usepackage{graphicx} 
\usepackage{epsfig} 
\usepackage{amsmath} 
\usepackage[latin1]{inputenc}
\usepackage[english]{babel}
\bibpunct{(}{)}{;}{a}{}{,}
\renewcommand{\arcsec}[1]{^{\prime\prime}\!\!\!#1\,}   
\renewcommand{\arcmin}[1]{^{\prime}\!\!\!#1\,}
\renewcommand{\mag}[1]{^{\rm m}\!\!\!#1\,}
\newcommand{\pc}{\ensuremath{\rm pc}}
\newcommand{\kpc}{\ensuremath{\rm kpc}}
\newcommand{\kms}{\ensuremath{\rm km~s^{-1}}}              
              
\newcommand{\AAA}{\ensuremath{\rm \AA}}              
\newcommand{\cmc}{\ensuremath{\rm cm^{-3}}}

\title{Spectral Variability of Romano's Star} 
\author{O. Maryeva\altaffilmark{1, 2} \and P. Abolmasov,\altaffilmark{3}}
\altaffiltext{1}{ Stavropol State University, Stavropol 355009, Russia }
\altaffiltext{2}{ Special Astrophysical Observatory, Nizhnij Arkhyz 369167, Russia }
\altaffiltext{3}{Sternberg Astronomical Institute, Moscow 119992, Russia }              
\shortauthor{Maryeva \& Abolmasov}
\shorttitle{Spectral Variability of V532}
\listofauthors{O.~Maryeva, P.~Abolmasov}
\indexauthor{Maryeva,~O.}
\indexauthor{Abolmasov,~P.}

\fulladdresses{
  \item Stavropol State University,
    Pushkina str., 1
    Stavropol, Russia 355009

  \item Special Astrophysical Observatory,
    Nizhnij Arkhyz,
    Zelenchukskij region,
    Karachai-Cirkassian Republic,
    Russia 369167
  \item Sternberg Astronomical Institute, Moscow State University
    Universitetsky pr., 13,
    Moscow, Russia 119992
}

\abstract{
We combine archival spectral observations of the LBV star V532 (Romano's star) 
together with the existing photometric data in the B band. 
Spectroscopic data cover 15 years of observations (from 1992 to 2007).
We show that the object in maximum of brightness behaves as an emission line 
supergiant while in minimum V532 moves along the sequence of late WN stars. 
In this sence, the object behaves similarly to the well-known Luminous
Blue Variable (LBV) stars AG Car and R127, but is somewhat hotter in
the minima. 
We identify about 100 spectral lines in the $3700\div 7300$\AAA\ wavelength range. For today, our spectroscopy
is the most comprehensive for this object. The velocity of the wind is
derived using HeI triplet lines ($360\pm30{\rm km~s^{-1}}$). 
Physical parameters of the nebula around V532 are estimated.
}
\addkeyword{stars: individual: Romano's star (M33)} 
\addkeyword{galaxies: individual: M33}
\addkeyword{stars: supergiants}
\addkeyword{stars: Wolf-Rayet}
\begin{document}
\maketitle

\section{Introduction} 

Luminous Blue Variables (LBVs) are a class of rare astrophysical objects 
introduced by \citet{conti84} and remain a subject of intense interest.
LBVs are broadly accepted as very massive and energetic stars emitting close to the 
Eddington limit, evolving from Of toward Wolf-Rayet stars. However, contemporary evolutionary 
models do not answer the question about the exact relations between LBV, nitrogen-rich Wolf-Rayet 
(WN) and hydrogen-rich WN (WNH) stars. 
 Numerical simulations show that the objects may pass the LBV 
evolutionary stage either before or after the WNH stage 
\citep{SmithConti}. 
Currently, only 35 LBV and LBV candidates are known in our
Galaxy  \citep{clark}. 
Studying LBVs in nearby galaxies is very important for 
understanding stellar evolution and the evolution of the interstellar
medium perturbed and contaminated by massive stars at various evolutionary stages.

The object V532 ($ \alpha = 01^{h} 35^{m} 09.^{s}71 $, $ \delta= +30^{o} 41\arcmin{}~57\arcsec{.}1$) 
is located in the outer spiral arm of the M33 galaxy. Giuliano Romano was the first to recover
its light curve  \citep{romano} 
 and to find irregular magnitude variations between 16$\mag{.}7$ and 18$\mag{.}1$.
Romano classified V532 as a variable of the Hubble-Sandage type by the
shape of the light curve and its colour index. 
\citet{HumphreysDavidson} 
 on the basis of its light variations classified the star as an LBV candidate.
Two maxima of brightness were detected during the last half of the
century \citep{kurtev}. 
The first was observed around 1970 and
the second in the early 1990s \citep{kurtev}. 
Long-term variability has an amplitude of about 1$\mag{\,}$ and seems to
be superimposed on an even stronger downward trend.
Short-timescale variability with amplitude $\sim$0$\mag{.}$5 was discovered in addition to the
longer-timescale variability
\citep{kurtev,olgaalla}. 
Such photometric behavior is typical for an LBV star.


First optical spectrum was obtained by T.~Szeifert 
\citep{szeifert} 
at the 3.5~m Calar Alto telescope in 1992.
In this spectrum, ``few metal lines are visible, although a late B spectral type is most 
likely (faint HeI)''.   
This spectrum, obtained with the TWIN spectrograph of the Calar Alto
telescope, is unique in being obtained in a profound flare state. It
is drastically different from the hot spectra observed in the
minima of brightness (see below). 

Second spectrum was obtained by O.~Sholukhova at the 6m telescope of the 
Special Astrophysical Observatory (SAO) of Russian Academy of Sciences 
(RAS) in 1994 \citep{olga}. 
 Another spectrum was
obtained at SAO with the Multi-Pupil Fiber Spectrograph (MPFS) in September 1998. 
This spectrum was classified as WN10--WN11 
\citet{fabrika}. 
\citet{polcaro}  
  estimated the bolometric absolute magnitude of
the object as $M_{bol} \simeq -10\mag{.}4$, using bolometric
correction ``of at least -3~mag'' and distance modulus
$m-M=24\mag{.}8$. 
They classify V532 as an LBV because the object fulfills all the
criteria of \citet{HumphreysDavidson}.  
  Using five spectra carried out in 2003 $\div$ 2006, 
\citet{viotti2006,viotti2007} 
 find anti-correlation between equivalent widths of the 
Wolf-Rayet blue bump at $4630\div 4686$\AAA\ and visual luminosity.

Comparing  the spectra  published by \citet{szeifert}, \citet{fabrika}
and  \citet{viotti2007}, 
we find that the object changes its spectral properties significantly. 

Here we combine the spectral observations (both new and
already published) with the light curve of the object
to trace the spectral variability of the star.
We describe the data and data reduction process in the next
section. Results are presented in section \ref{sec:res} and discussed
in section \ref{sec:disc}.   
In section \ref{sec:con}, we present the conclusions.

\begin{table*}[p]\centering
\caption{Observational log for MPFS data. S/N is signal-to-noise ratio
  per resolution element.}\label{tab:obstabmpfs}

\begin{tabular}{lccccc}

\toprule
Date         & Exposure  & Seeing, $ \arcsec$ &  S/N & Spectral      & Spectral  \\
             &   time, s &                    &      & standard star & range, \AAA    \\
\midrule
05. 10. 2002 &  900      & 3.8                &   16   & BD25d4655    &  4250-6700   \\

13. 11. 2004 &  4200     & 1.5                &   20  & HZ44         &   4000-7000   \\

17. 01. 2005 &  3600    & 1.5                &    18  &  G248        &  4000-7000     \\

\end{tabular}

\end{table*}

\section{Observations and Data Reduction}\label{sec:obs}

In this work, we use archival data from the 6m SAO telescope (available
via ASPID database, \url{http://alcor.sao.ru/db/aspid/}) and the SUBARU 
telescope, which is operated by the National Astronomical Observatory of Japan.   
The 6m telescope data were obtained with the Multi Pupil Fiber Spectrograph (MPFS)
\citep{mpfs} 
 and with the SCORPIO multi-mode focal reducer in the long-slit mode   
\citep{scorpio}. 
The data from SUBARU were obtained  with the Faint Object Camera (FOCAS) 
\citep{focas} 
 in the Cassegrain focus. 
 Uncertainties are everywhere dominated by statistical Poissonian
 noise (readout noise and round-off errors are significantly smaller).

\subsection{MPFS Data}

MPFS obtains simultaneously the spectra from 240 ($16\times 15$) or 256 ($16\times 16$) 
spatial sampling elements arranged in the form of a rectangular array of square lenses. 
In 2002 and 2004-2005, the data were obtained in the 240-element
and 256-element configurations, respectively. 
Spatial sampling of 1$\arcsec{\,}\times$1$\arcsec{\,}$ was used. 
Light from individual sampling elements is collected by microlenses
and transmitted by means of optical fibers reorganized in the form of
a pseudo-slit toward the spectral camera. 
Grating \#~4 (600 \AAA\,$\rm pix^{-1}$) providing spectral resolution
of about 6~\AAA\ was used for all the observations.
Detectors CCD TK 1024 ($1024\times 1024$ pixels) and
EEV42-40 ($2048\times 2048$ pixels) were used in 2002 and 2004-2005, correspondingly. 
Sky background spectrum at the distance of 4$\arcmin{\,}$ away from the
object is taken simultaneously with the object by 17 
(for 16$\times$16 field size) or 16 (for 16$\times$15 configuration)
additional fibres. 
We summarize the relevant information on the MPFS data in
table~\ref{tab:obstabmpfs}. 

Data reduction system was written in IDL environment and makes use of procedures 
written by V. Afanasiev, A. Moiseev and P. Abolmasov. 
Reduction process consists of the standard steps for panoramic data
reduction (see for example \citet{sanchez}):
bias subtraction, flat-fielding, removal of cosmic-ray 
hits, extraction of the individual spectra from the frames 
 and their wavelength calibration using a spectrum of an He-Ne-Ar lamp. 
At every wavelength, we calculate the median sky level using 
the offset fibres and subtract from the spectra of the field. 
Spectra of spectrophotometric standard stars were used 
for absolute flux calibration. 

Three emission-line spectra obtained with MPFS in 2002-2005 were
used. We estimate signal-to-noise ratio in continuum as 10$\div$30 per
resolution element (see table~\ref{tab:obstabmpfs} for details).
Integral spectra were extracted in an annular aperture $2\arcsec{\,}$
in radius. Up to the instrumental resolution, the 
object point-like. Note that, unlike long-slit spectrographs, MPFS is
free from slit losses at absolute flux calibration is much more
reliable. 

\subsection{SCORPIO Data}\label{sec:scorpio}

Observational data from SCORPIO are summarized in table~\ref{tab:obstabscor}.
All the spectra were reduced using {\tt ScoRe} package for long-slit data
reduction, written in IDL
(available at \url{http://narod.ru/disk/5238009000/Score_v1.2.tar.html}).
CCD frames were de-biased, cosmic particle hits were removed from all
types of exposures except bias. Save for the single 10 minute long
exposure obtained in February 2005, several object exposures are
present for every day that simplifies the removal of cosmic hits. 
Then we divide object and spectral standard exposures over normalized mean flat-field frame.
After wavelength calibration using He-Ar-Ne lamps and night sky OI
($\lambda 5577$ or $\lambda 6300$ depending on grism) emission lines,
the CCD data were flux-calibrated using standard stars from \citet{oke} spectrophotometric standard
list (see table~\ref{tab:obstabscor}). 
Spectra were extracted by fitting the profiles of slices across
dispersion with  Gaussian function.  

\begin{figure}
\centering
\epsfig{file=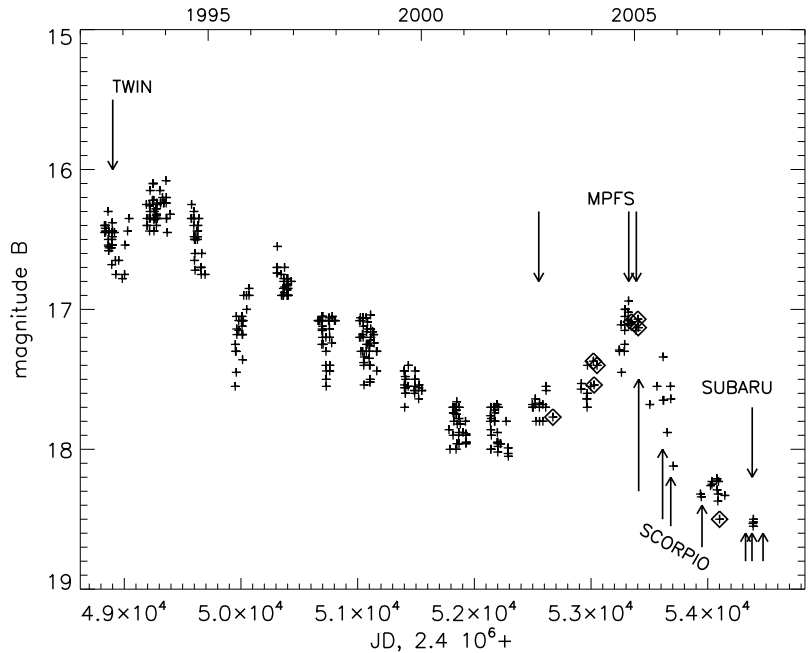,width =\linewidth}
\caption{B-band light curve of V532. 
Arrows indicate when spectroscopic observations were made. 
 Data from \citep{viotti2007} are shown by diamonds.
}
\label{fig:lightcurve}
\end{figure}

In total, eight spectra were obtained with SCORPIO in 2005-2008 (dates
of observations are shown
by upward arrows in figure \ref{fig:lightcurve}).
Spectral resolution is 10~\AAA (for VPHG550G grism) and 5~\AAA (for
VPHG1200G and VPHG1200R grisms), signal-to-noise ratios in continuum
vary from about 10 to about 50 for larger total exposures.  

\begin{table*}\centering
\caption{Observational log for the SCORPIO data. $\alpha$ is seeing, PA is position angle, S/N is signal-to-noise ratio.}\label{tab:obstabscor}
\begin{changemargin}{-0.5cm}{-0.5cm}
\setlength{\tabcolsep}{1\tabcolsep}\tablecols{8}
\begin{tabular}{lcccccccc}
\toprule
Date      & Exposure & Grism & Spectral     & $\delta \lambda ,$ &   S/N    &$\alpha,~\arcsec$  & Spectral      & PA,$^{o}$\\
          &   time, s&       &    range,~\AAA &    ~\AAA         &          &                   & standard &          \\
          &          &       &              &                    &          &                   & star     &           \\     
\midrule

6. 02. 2005 &  600     &       &              &                    &   8      & 1.7               & G248          & -136  \\

\cmidrule(l){1-2}
\cmidrule(l){6-8}

30. 08. 2005& 1200    & VPHG550G & 3500-7200  &  10                &   30    &1.9                & G191-B2B      & 210  \\

\cmidrule(l){1-2}
\cmidrule(l){6-8}

8. 11. 2005 & 3300    &          &            &                    &   45    & 1.9               &  BD25d4655    & 145  \\

\midrule

3. 08. 2006 & 1500    &          &            &                      &    20    &2                &  GD248        &  200 \\
\cmidrule(l){1-2}
\cmidrule(l){6-8}

10. 08. 2007& 1800    & VPHG1200G&  4000-5700 & 5                    &   24     &2                 & BD33d2642     & 252  \\

\cmidrule(l){1-2}
\cmidrule(l){6-8}

5. 10. 2007 & 2700    &          &            &                     &     40  & 1.1                &  BD25d4655    & -141  \\
\midrule
 8. 01. 2008&  1800   & VPHG1200R&  5700-7500 & 5                    &    20    &2.1                &  BD25d4655    &  48   \\
\midrule
 10. 01. 2008&  1800   & VPHG1200G&  4000-5700 & 5                    &   20    & 1.4                & BD25d4655       &  18   \\
\end{tabular}
\end{changemargin}
\end{table*}

\subsection{Data from SUBARU}

One exposure 1200s in length was obtained with the SUBARU telescope in October 2007. 
VPHG450 grism was used providing the spectral range of 3750-5250~\AAA. 
Slit width of 0$\arcsec{.}5$ implies spectral resolution of about 1.7~\AAA\ 
(for today, it is the best resolution achieved for this object) in the extracted spectrum.
Data were reduced using IDL-based software. 
The CCD frames were bias-subtracted and  flat-fielded. We used the Th-Ar arc spectrum for wavelength calibration.
Star BD40d4032 from SUBARU spectrophotometric standard list  was used to calibrate the stellar spectra. 
We extract the spectrum in the same way we did it for SCORPIO (see above section \ref{sec:scorpio})

If we degrade the spectral resolution to 5\AAA, the spectrum
becomes practically 
identical to the spectrum obtained with SCORPIO on the same date. 
Though the spectral shapes are similar, the spectra differ in normalization 
(about a factor of three) connected to slit losses.

\subsection{Photometric Data}

The light curve (see figure~\ref{fig:lightcurve}) was provided by
Vitalij Goranskij and consists of the CCD data obtained by Goranskij and Zharova with the 1m SAO telescope and two 
instruments of the Crimean laboratory of the Sternberg Astronomical Institute (SAI) and photographic plates 
from the SAI collection. CCD data were reduced with MaxIm DL software
(\url{http://www.cyanogen.com/maxim_main.php}). The joint light curve
will be published by \citet{ZGphoto} in a separate paper, all the
details of the data reduction process will be given there. 
Photographic B-band magnitudes were identified with the B-band
magnitudes obtained by CCD observations.  
CCD data uncertainties are of the order $0\mag{.}05$, plate data have larger errors 
depending on the source brightness, usually of the order $0.1- 0.2\mag{\,}$.

The joint curve contains the data of \citet{viotti2007} but
complements it with a much larger photometric material and
primarily allows to trace the recent evolution of the object.

\section{Results}\label{sec:res}

\subsection{Spectral Evolution}

\begin{table*}
\centering
\tablecols{6}
\setlength{\tabnotewidth}{1\textwidth}
\caption{List of emission lines detected in the FOCAS and
  SCORPIO (January 2008) spectra of V532. For
  higher signal-to-noise ratios, equivalent widths (EW) are given. For
lines with P~Cyg profiles, we give both emission and absorption
component EWs. }\label{tab:linelist}
  \begin{changemargin}{-1cm}{1cm}
\setlength{\tabcolsep}{0.8\tabcolsep} \tablecols{8}
\begin{tabular}{lccc@{\hspace{6\tabcolsep}}lccc}
\toprule
$\lambda$, \AAA\ &    Ion    & EW       & EW         &   $\lambda$, \AAA\ &    Ion    & EW       & EW        \\
         &           & emission, & absorption, &            &           & emission & absorption \\
         &           &   \AAA    &   \AAA      &            &           &   \AAA    &    \AAA       \\

\midrule
 3770.60          &  H11+HeII             &    &                   &   4613.90  &  NII     &    &  \\
 3797.90          &  H10+HeII             &    &                   &   4621.40  &  NII     &    &  \\
 3819.76          &  HeI                  &    &                   &   4630.54  &  NII     &    &  \\
 3835.39          &  H9+HeII              &    &                   &   4634.00  &  NIII    & $6.2\pm 1.0$   &  \\
 3871.82          &  HeI                  &    &                   &   4640.64  &  NIII    &  $7.4\pm 0.5$  &  \\
 3889.05          &  H8+HeI               &$11 \pm 1$& $3.7\pm1.3$ &   4643.09  &  NII     &    &  \\
 3964.73          &  HeI                  &    &                   &   4650.16  &  C III   & $5.0\pm 0.5$    &  \\
 3970.08          &  H$\epsilon$+HeII     &    &                   &   4658.10  &  [FeIII] & $2.5\pm 1.0$   &  \\
 3994.99          &  NII                  &    &                   &   4685.81  &  HeII    & $15.2\pm0.2$   &  \\
 4009.00          &  HeI                  &    &                   &   4701.50  &  [FeIII] &  $2.9 \pm 1.0$  &  \\
 4025.60          &  HeI+HeII             &2.5$\pm$0.5& 1.2$\pm$0.2&   4713.26  &  HeI     &  $1.6 \pm 0.2$  &  \\
 4088.90          &  SiIV                 &    &                   &   4861.33  &  H$\beta$&  $22.5 \pm 1.5 $ &  \\
 4097.31          &  N III                &    &                   &   4921.93  &  HeI     &  4.3$\pm$0.2   &  \\
 4101.74          &  H$\delta$+HeII       & 7.6$\pm$1.0   &        &   4958.91  &  [O III] &1.3$\pm$0.2    &  \\
 4103.40          &   NIII                &    &                   &   5006.84  &  [O III] &3.9$\pm$0.6    &  \\
 4116.10          &  SiIV                 &    &                   &   5015.67  &  HeI     &2.6$\pm$0.3    &  \\
 4120.99          &  HeI                  &    &                   &   5411.50  &  HeII    &6.2$\pm$0.2    &  \\
 4143.76          &  HeI                  &    &                   &   5666.60  &  NII     &    &  \\
 4199.80          &  HeII                 &    &                   &   5676.02  &  NII     &    &  \\
 4236.93          &  NII                  &    &                   &   5679.56  &  NII     &    &  \\
 4241.79          &  NII                  &    &                   &   5686.21  &  NII     &    &  \\
 4241.79          &  NII                  &    &                   &   5710.76  &   NII    &    &  \\
 4340.47          &   H$\gamma$+HeII      &8.0$\pm$0.3&            &   5875.79  &  HeI     & 25.5$\pm$0.7   &  \\
 4387.93          &  HeI                  &1.4$\pm$1.0    &        &   6548.00  &  [N II]  &    &  \\
 4471.69          &  HeI                  & 5.0$\pm$0.3 & 1.7$\pm$0.2&   6562.82  &  H$\alpha$+HeII  &107$\pm$3    &  \\
 4481.13          &  MgII                 &    &                   &   6583.00  &  [N II]  &    &  \\
 4510.90          &  NIII                 &    &                   &   6678.15  &  HeI     &18 $\pm$3   &  \\
 4514.90          &  NIII                 &    &                   &   6683.20  &  HeII    &    &  \\
 4518.20          &  NIII                 &    &                   &   7065.44  &  HeI     &18 $\pm$3     &  \\
 4523.60          &  NIII                 &    &                   &   7135.73  &  ArIII   & $3.2\pm 0.3$   &  \\
 4530.80          &  NIII                 &    &                   &   7281.35  &  HeI     &    &  \\
 4534.60          &  NIII                 &    &                   &            &          &    &  \\
 4541.60          &  HeII                 &    &                   &            &          &    &  \\
 4547.30          &  NIII                 &    &                   &            &          &    &  \\
 4601.50          &  NII                  &    &                   &            &          &    &  \\
 4607.20          &  NII                  &    &                   &            &          &    &  \\

\end{tabular}
   \end{changemargin}
\end{table*}

\begin{figure}
\centering
\epsfig{file=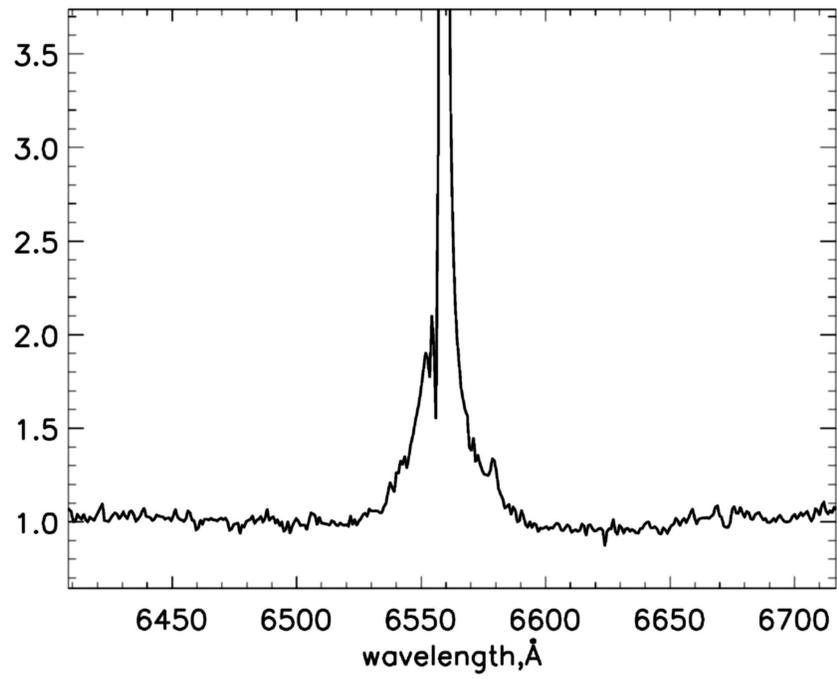,width = \textwidth}
\caption{H$\alpha$ line profile in the Calar Alto spectrum
  obtained in 1992. The spectrum is normalized by the local continuum level.
}
\label{fig:twinhalpha}
\end{figure}

 The first optical spectrum of V532  was obtained during the rise 
of brightness, when the object was 16$\mag{.}$4 in the V band.  This spectrum covers two spectral ranges
$\lambda \lambda 4400-5000$ and $\lambda \lambda 5800-6750$, as it was
obtained with the two-channel TWIN spectrograph (a comprehensive
description is present at \url{http://www.caha.es/CAHA/Instruments/TWIN/HTML/twin.html}).
In this spectrum, two strong lines are present, $H\alpha$ (shown in figure~\ref{fig:twinhalpha}) and $H\beta$, 
having complex profiles consisting of a narrow line with P~Cyg profile and broad wings.
FWHM (Full Width at Half Maximum intensity) of the wings is $\sim 15 \AAA$ for H$\beta$ and $\sim 20 \AAA$ for H$\alpha$. 
Such wings were first found for LBVs in the spectrum of P Cygni by \citet{bernat}. 
These wings are explained by scattering of line photons by free electrons in the stellar wind. 
Similar line profiles were observed for the LBV stars R127 and AG~Car during
the initial rise to maxima \citep{stahl1983,stahl2001}. 

Emission lines of HeI$\lambda 5876$ and SiII$\lambda5957.612, 
5978.970, 6347.091, 6371.359 $ are also present in the red spectrum. 
\citet{szeifert} 
 mentions these SiII lines as ``weak metal emissions''.

\begin{figure*}
\centering
\epsfig{file=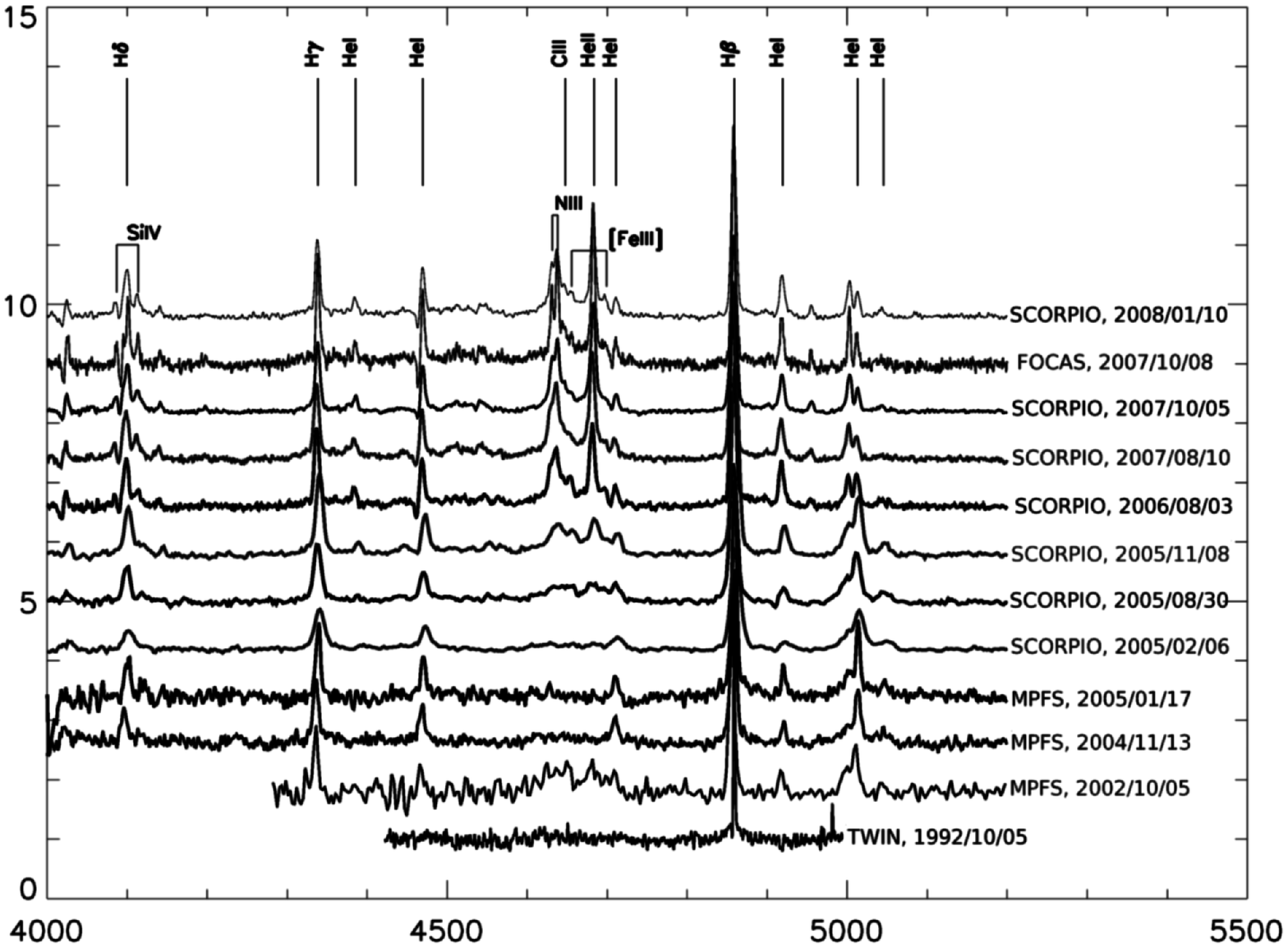,width=1.\textwidth}
\caption{Optical spectrum evolution in the blue spectral range (4000$\div$5200\AAA).
Spectra are normalized by the local continuum level and vertically shifted
for clarity.}
\label{fig:spevolution}
\end{figure*}

Figure~\ref{fig:spevolution} shows all the spectra of V532 in the
blue range (4000-5500\AAA) analysed in this article, obtained 
at different spectral resolutions. 
All the spectra are normalised to continuum level in a uniform
way. In order to get a reasonable normalisation for our spectra, we chose several 
wavelength intervals practically free from Wolf-Rayet emissions (4250$\div$4270, 5100$\div$5200, 
5520$\div$5620, 5750$\div$5800 and 6950$\div$6970$\AAA$)   
and reconstructed the continuum using a second-order polynomial. 
Below, for spectral classification we use characteristic
equivalent width ratios, that makes our results practically independent
of spectral resolution.

The spectral appearance of the spectra of V532 obtained in 2002-2008
resembles that of late WN stars. 
It allows to apply a classification scheme used for WN stars, bearing
in mind that abundances of individual elements may differ, but
physical conditions are similar.
 All the spectra obtained between 2002 and 2008 were 
classified using the classification of \citet{smithprinja} for WN6-11 stars based 
primarily on relative strengths of NV~$\lambda\lambda4604-20$, NIV 
$\lambda4058$, NIII $\lambda\lambda 4634-41$ and NII $\lambda3995$
emission lines. 
 The method has low dependence on elemental abundances, though only helium and
nitrogen lines (preferably, ratios of the lines of one element) are used. 
The results of spectral classification are given in table
\ref{tab:spsmith}.

\begin{table*}
\centering
\tablecols{6}
\setlength{\tabnotewidth}{1\textwidth}
\caption{ Spectral classes identified via the scheme of Smith,
  Crowther and Prinja~(1994) }\label{tab:spsmith}

\begin{tabular}{lllc@{\hspace{4\tabcolsep}}clll} 
\toprule

 B, mag&  Spectral &  Date     && & B, mag&  Spectral &  Date  \\

       &  subtype  &         &&   &       &  subtype  &        \\
\cmidrule(l){1-3} 
\cmidrule(r){6-8}
17.5    & WN10.5 & 2002/10/05   & &&         17.6 & WN9  & 2005/11/08\\
16.9    & WN11   & 2004/11/13   & &&         18.3 & WN8  & 2006/08/03\\ 
17.1    & WN11   &  2005/01/17  & &&         18.4 & WN8 & 2007/08/10\\ 
17.15   & WN11   & 2005/02/06   & &&         18.5 & WN8 &   2007/10/05-08\\  
17.3    & WN10   & 2005/08/30   & &&              & WN8  &2008/01/08-10\\
\bottomrule
\end{tabular}

\end{table*}


 Light curve exhibits a local maximum in 2004 and early 2005. 
 In the spectra obtained in this period, one may observe strong emission lines 
of hydrogen and neutral helium. The spectral appearance of V532 shows
 strong similarities with a spectrum of a WN11 star. HeI~$\lambda 4713$ line is stronger
 than the NII blend, and  HeII$\lambda4686$ is absent. 
In the spectrum obtained in 2002, the NII~$\lambda$3995 line is
outside the spectral range, therefore we carry out classification
comparing NIII~$\lambda\lambda4634-41$ and
NII~$\lambda\lambda4601-43$ blends having similar intensity to the
NII~$\lambda$3995 emission. NIII~$\lambda\lambda4634-41$ lines
appear, but are weaker than NII~$\lambda\lambda4601-43$. 
HeII$\lambda4686$ line is approximately as bright as the NII~$\lambda\lambda 4634-41$
 emission, hence we classify the object as WN10.5. 

Starting from the middle of 2005, Romano's star weakens in the optical range, its 
visible magnitude falls from 17 to 18$\mag{.}$8. During this period
its spectrum evolves from WN10 (August 2005) through WN9 (November
2005) to WN8 (August 2007) by three spectral sub-classes. 
In October 2007, HeII~$\lambda$4686 line is stronger than
NIII~$\lambda\lambda 4634-41$. However, NIV lines do not appear in the
spectrum. We also classify the spectrum as WN8.

\begin{figure*}
\centering
\epsfig{file=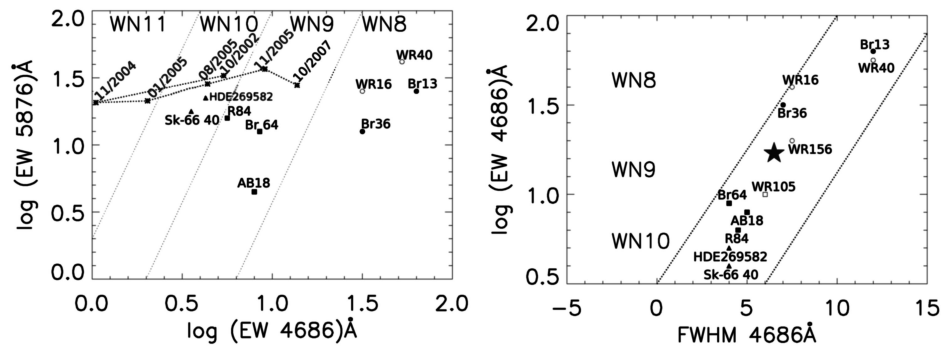, width =1.1\linewidth}
\caption{Left panel: V532 location on equivalent width 
  diagram of ${\rm HeI}\lambda 5876$ versus ${\rm HeII} \lambda 4686$
  for different dates of observation. The points are connected with
  dotted lines in chronological order.
Right panel: V532 location in October 2007 (FOCAS data) on
equivalent width versus FWHM diagram for the ${\rm HeII}\lambda 4686$ line. In
both graphs, known Galactic (open symbols)
  and LMC (filled) WN stars are shown for comparison:
  WN8 by circles, WN9 by squares, and WN10 by triangles. Data on these
  objects were taken from \citet{smith}. 
}
\label{fig:spclass}
\end{figure*}

V532 evolution on the equivalent width diagram of HeI~$\lambda 5876$
 versus HeII~$\lambda 4686$ is shown in the left panel of figure~\ref{fig:spclass}. 
 Locations of well-proven Galactic and Large Magellanic Cloud (LMC) Wolf-Rayet stars 
R84 (WN9), Sk-60~40 (WN10), LBV HDE~269582 (WN10) and some other objects are shown for comparison with our object.
The figure shows that in the period between 2002 and 2005, when the
star brightened by about one magnitude, its spectrum changed from WN9.5
to WN11. Since 2005, the spectrum changes smoothly according to the
sequence established by \citet{shara}. 

We used a quantitative chemistry-independent criterion based on the
FWHM of the HeII~$\lambda 4686$ line for alternative spectral classification. 
In Fig.\ref{fig:spclass}, we show the location of V532 on the diagram
of the equivalent width of HeII~$\lambda 4686$ versus FWHM of this
line. We have measured equivalent width and FWHM of the HeII line in
the FOCAS spectrum. For this, we approximate Wolf-Rayet blue bump by 7 Gaussians. 
The position of V532 is fully consistent with its Galactic and LMC WN9
analogues. Spectral class defined from the diagram is consistent with
that determined from the relative strengths of NII, NIII and
HeI with accuracy about one subclass. 

Table~\ref{tab:linelist} shows the lines detected in the spectra obtained
with FOCAS (October 2007) and with SCORPIO (January 2008, with grisms 1200G and 1200R). 
Three spectra are used in order to cover the maximal wavelength range at 
maximal possible spectral resolution. The spectrum of V532 does not change 
noticeably from October 2007 to January 2008.
Equivalent widths of the principal lines (absorptions,
emissions and P~Cyg) are given with errors of approximation. 
Uncertainties due to the choice of the continuum level are smaller than the
errors of approximation. Lines with P Cygni profiles were resolved by
fitting with the models described in the next subsection. 

\subsection{Line Profiles and Terminal Wind Velocity} 

\begin{figure*}
\centering
\epsfig{file=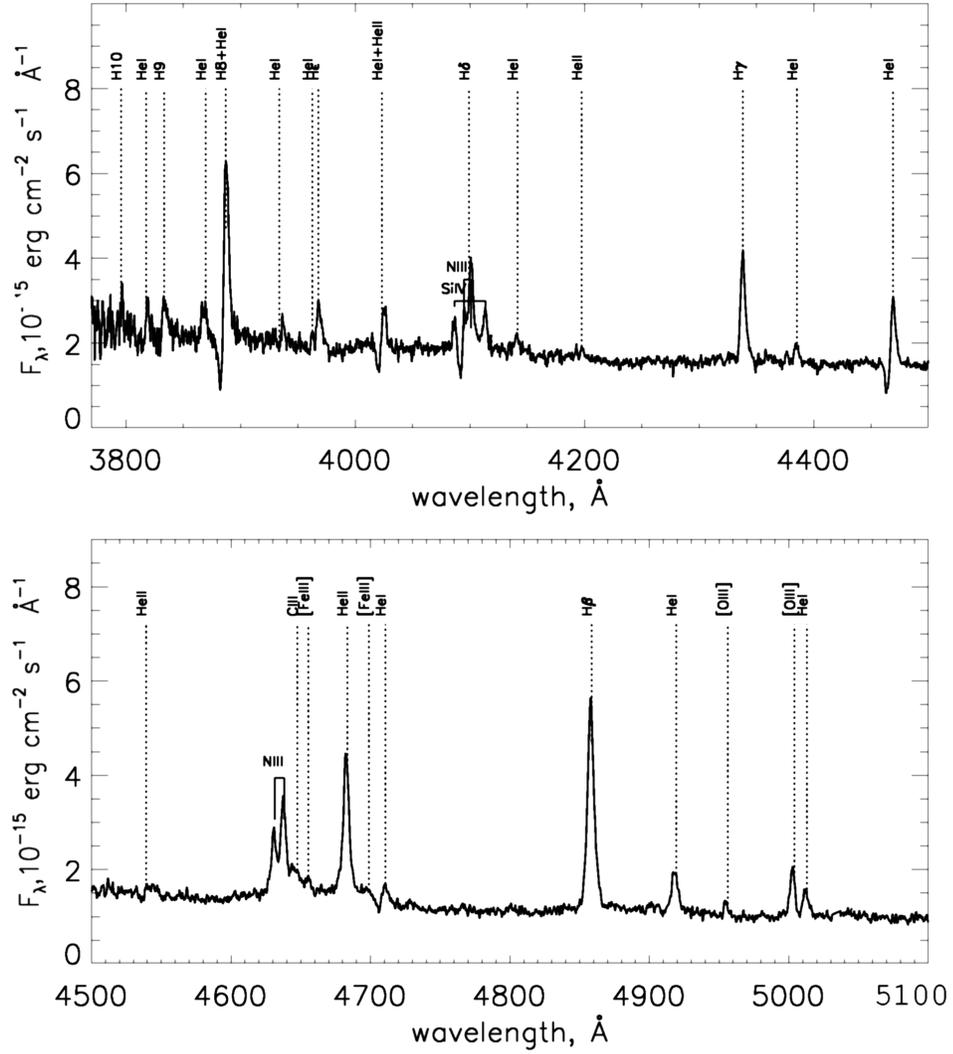,width =1.1\linewidth}
\caption{Optical spectrum of V532 obtained with FOCAS in the range 3800-5100\AAA\AAA.}
\label{fig:subaruspect}
\end{figure*}

    In figure \ref{fig:subaruspect} we show the FOCAS spectrum in the optical
blue range $3800-5100\AAA$. The Wolf-Rayet blue bump 
(consisting primarily of NIII$\lambda\lambda 4634, 4640$, CIII$\lambda 4650$,
HeII$\lambda 4686$ and HeI$\lambda 4713$) is clearly seen in this spectrum.
 We suppose that the lines near 4658 and 4701 \AAA\ are nebular lines of [FeIII].
They can not belong to CIV since CIV~$ \lambda \lambda5812, 5801$ lines 
are not present in the SCORPIO spectrum obtained simultaneously in October 2007.

\begin{figure*}
\centering
\epsfig{file=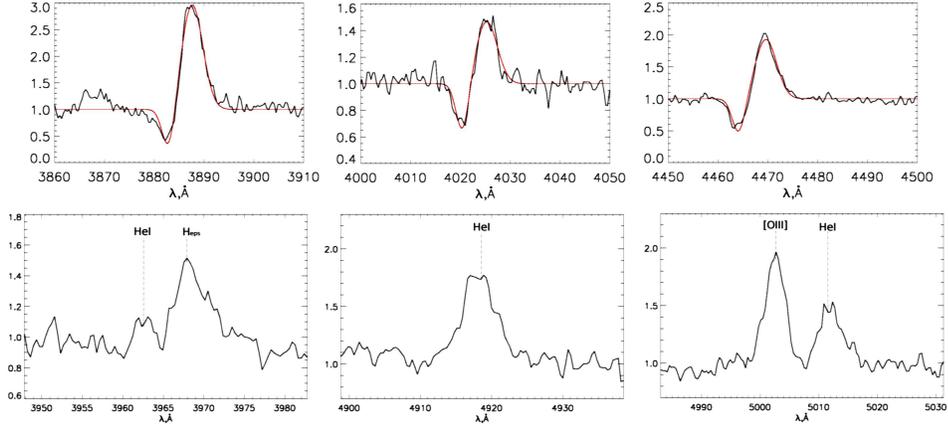,width =1.1\linewidth}
\caption{Top panel: profiles of triplet HeI$\lambda3889,4025,4471$
  lines (from left to right). Two-component model fits are shown (see text for details).
Bottom panel: profiles of singlet HeI$\lambda3965,4922,5016$ lines. FOCAS data.}
\label{fig:singlet}
\end{figure*}

   Analyzing the FOCAS spectrum of V532, we found that triplet and singlet 
lines of HeI have different profiles (see figure~\ref{fig:singlet}). 
Triplet lines of HeI ($\lambda3889,4025,4471$) show strong P~Cyg profiles, 
while singlet lines ($\lambda 3965, 4922, 5016$) have flat-topped profiles. 
Widths of these lines correspond to velocity span of about $100~\rm km~s^{-1}$. 

We used P Cyg profiles of triplet HeI lines to estimate the terminal wind velocity $v_{\infty}$. 
Line profiles were fitted with a sum of two Gaussians
(one representing the absorption, the other the emission component).  
We suppose that the widths of emission and absorption components 
are equal to the instrumental profile width of $1.8$\AAA\ which is almost independent of wavelength. 
The instrumental profile width is determined using wavelength calibration 
spectrum lines of similar wavelengths.
Terminal wind velocity is estimated by the velocity shift between the Gaussian centers.   
Even for the emission component of HeI lines, we do not detect any
Doppler broadening. This may mean that the profiles are not true 
P~Cyg but wind blueshifted absorptions plus nebular emissions. This  
possibility does not however alter the wind velocity
estimates. 

Wind velocities for all the HeI lines with P~Cyg profiles  
are equal within the statistical errors. 
The mean wind velocity for the  three triplet lines is $360\pm 30~{\rm km~s^{-1}}$. 
This value is consistent with the terminal wind velocities for late
WN \citep{Crowther1995,smith} 
that we give in table \ref{tab:windvelocity} for comparison. 

\bigskip

It is interesting to compare these profiles with the 
HeII$\lambda5412$ line in the SCORPIO spectrum obtained in October
2007. The spectrum has very high signal-to-noise ratio that allows
reasonable terminal wind velocity estimates in spite of the
significantly worse spectral resolution. 
The line has a classical low-optical-depth 
P~Cyg profile: broad emission and blueshifted narrow absorption. 
We approximate this line by a classical P~Cyg model 
profile (flat-topped emission with a narrow blueshifted absorption
component) convolved with instrumental point spread function that we
consider Gaussian 
(the instrumental profile width is $\sim 5.5$\AAA). 
Similar line profiles are produced by optically thin envelopes
expanding at a constant velocity. 
Its fitting yields the wind speed of $200\pm 15~{\rm km~s^{-1}}$.
The expansion velocity indicates that the line is formed in
hotter inner parts of the wind that move with lower outward
velocities.
Among Pickering emissions, we also detect 
$\lambda 4541.6$ and $\lambda4199.8$ having similar profiles in the FOCAS spectrum, but we lack
signal-to-noise to obtain reliable expansion velocity estimated for these lines. 

\subsection{Nebular Lines}\label{sec:nebula}

The lines at 4959 and 5007 \AAA\ are clearly identified as [OIII] emissions, 
that is consistent with their flux ratio equal to 3 and FWHMs equal within the measurement errors.
In the cooler spectra of 2003 \citep{polcaro}, 
nebular lines of [OIII] 
are overridden by NII 4994-5005 emissions. However, in 2007, NII lines 
are absent, but the [OIII]$\lambda$4959,5007 doublet is clearly seen.

Nebular lines of [OIII]$\lambda\lambda 4959, 5007$, 
[NII]~$\lambda\lambda 6548-83$ are present in all the spectra analysed by us 
while [SII]~$\lambda6717-31$  doublet is not. 
Therefore, we can estimate the electron density of the surrounding nebula. Electron
density is between $10^{3.6} \rm cm^{-3}$ and $10^{4.9} \rm cm^{-3}$.
The former value corresponds to the critical density of the [SII]
doublet, the second to that of [NII]$\lambda\lambda 6548-83$ lines (see for example \citet{Osterbrock}).

All the nebular lines present in the FOCAS spectrum as well as the
[ArIII]$\lambda7135.73 $ emission  in SCORPIO data show flat-topped or
two-peaked profiles. The widths of emission line cores are of the
order 100 \kms. 
Once we know the density of the emitting gas, we may estimate the size
of the Str\"{o}mgren region and the mass of the nebula surrounding
V532. The former may be found as follows (see for example \citet{leng}):

$$ R_{s}\simeq \left(\frac{3S}{4\pi \alpha n_e^2}\right)^{1/3} $$

where $\alpha$ is Case B recombination coefficient for hydrogen,
$\alpha \simeq 2.6 \cdot 10^{-13} \left(\frac{10^4}{T}\right)^{0.85}
\rm cm^3  \rm s^{-1}$, $n_e$ is electron density, and $S$ is the
number of hydrogen-ionizing quanta production rate. For O9.5Ia stars,
probably similar to the object in mass and luminosity,
$S$ is estimated as $10^{49.17}$~s$^{-1}$ \citep{Osterbrock}. 
One may estimate the size of a homogeneous nebula around V532 as:

\begin{equation}\label{E:rstrom}
R_{s}\simeq 0.1  \cdot \left(\frac{S}{10^{49}{\rm
    s^{-1}}}\right)^{1/3} \left(\frac{n_e}{10^{4} \cmc}\right)^{-2/3} \left(\frac{T}{10^4\rm K}\right)^{0.28}~~ \pc
\end{equation}

Note that possible inhomogeneity has little effect on the linear size: 
for a gas filling only some part of the nebular volume $f$,
$R_{s}$ should scale with filling factor as $\propto f^{-1/3}$. However, 
sizes of about $1 \pc$ are plausible. Nebular mass may be trivially inferred as:

\begin{equation}\label{E:mstrom}
M \simeq \rho \cdot \frac{4 \pi}{3} R_{s}^3 \simeq 0.17 \cdot \left(\frac{S}{10^{49}{\rm
    s^{-1}}}\right) \left(\frac{n_e}{10^{4} \cmc}\right)^{-1} 
\left(\frac{T}{10^4\rm K}\right)^{0.84} \cdot M_{\odot} 
\end{equation}

Estimated size and the ionised mass of the nebula are consistent with 
these for ejecta of LBV stars (see \citet{smithprinja} 
and references therein) by order of magnitude. It is possible that the
total mass lost by the object is higher, about several Solar masses,
but we observe only the ionized part of the envelope. The emitting gas
was probably ejected during one or several outburst events at wind
velocities about 100\kms.

\section{Discussion}\label{sec:disc}

In table \ref{tab:windvelocity}, terminal wind velocities for the WN9-WN11 stars in the
Galaxy, LMC and M33 and LBV star AG Car (during minimum) are
given. These velocities are also estimated through the optical HeI lines.
Because wind acceleration is tightly connected to metallicity
\citep{massey,puls}, 
we also give oxygen abundances (12 + lg O/H) for selected HII regions
adjacent to MCA1-B, B517 and BE381. They may be used as a measure for
initial object metallicities. 

V532 is located at a distance of about $17\arcmin{\,}$\ from the centre 
of M33. Taking the central oxygen abundance value of 
$12+\log{O/H}=8.36\pm0.04$ and radial gradient of 
$-0.027\pm0.012~\rm{dex}~\kpc^{-1}$  
 \citep{OHabundance}, 
we may estimate the primordial oxygen abundance for V532 as $8.26\pm0.08$. 
Ambient metallicity is thus similar to that for the 30~Dor 
star-forming region. 
Our estimate for the terminal wind velocity of V532 is fairly 
consistent with this for other late WN stars in similar environment.

\begin{table*}
\centering
\tablecols{6}
\setlength{\tabnotewidth}{1\textwidth}
\caption{Wind velocities estimated through optical HeI
    lines and ambient oxygen abundances.}\label{tab:windvelocity}

\begin{tabular}{lccccc} 
\toprule
                       
Star                                                              &
Galaxy   & WN      & ${\rm v_{\infty}}$ & $12+\log{\rm O/H}$   & HII\\
                                                                  &             & subtype & ${\rm km~s^{-1}}$  &  abundance       & region    \\
                                                                  &             &         &                    &                  &      \\
\midrule
V532                                                              &     M33     &      9  &      360           & $8.26\pm0.08$    &       \\
(during minimum                                                   &             &         &                    &                  &       \\
epoch 2007)                                                       &             &         &                    &                  &       \\

MCA1-B\tabnotemark{a}
                                                                  &     M33     &      9  &      420           & $8.315\pm0.061 $ & NGC588\tabnotemark{e} 
\\
B517 \tabnotemark{b} 
                                                                  &     M33     &      11 &      275           &  $8.334\pm0.083$ & MA2\tabnotemark{e}  \\
Sk-66$^{\circ} 40$\tabnotemark{a}
                                                                  &     LMC     &      10 &      300           &                  &      \\
R84\tabnotemark{a}  
                                                                  &     LMC     &      9  &      400           &                  &       \\
BE381\tabnotemark{c}   
                                                                  &     LMC     &      9  &      280           &   8.37           &30~Dor\tabnotemark{f}\\  
WR105 \tabnotemark{a}
                                                                  &     MW      &      9  &      700           &                  &         \\

AG Car                                                            &             &         &                    &                  &          \\
 (during minimum                                                  &      MW     &  11     &  300               &                  &       \\
 epoch 1985-1990) \tabnotemark{d}
                                                                  &             &         &                    &                  &       \\
        
AG Car                                                            &             &         &                    &                  &           \\
 (during minimum                                                  &    MW       &  11     &       200          &                  &       \\
  epoch 2002)  \tabnotemark{d}
                                                                  &             &         &                    &                  &     \\  
\bottomrule
\tabnotetext{a}{\citet{smith}} 
\tabnotetext{b}{\citet{b517}}
\tabnotetext{c}{\citet{Crowther1995}}
\tabnotetext{d}{\citet{AGcar}}
\tabnotetext{e}{\citet{OHabundance}}
\tabnotetext{f}{\citet{OHabundanceDor}}
\end{tabular}

\end{table*}

Wolf-Rayet stars are believed to be more evolved objects than
Ofpe and Be supergiants. Information on elementary abundances may 
cast some light upon this issue. However, to recover reliable
abundance estimates, one needs sophisticated modelling of both the
moving atmosphere of the star and its structure. Here, we restrict ourselves
only to some semi-qualitative estimates for the H/He abundance ratio.
We compare the equivalent width ratios of H$\beta$ to helium lines in the spectra obtained in 2007 
to those for several WN9 (taken from 
\citet{Crowther1995}) 
 to estimate the relative abundances of hydrogen and helium.
The equivalent width ratios are given in table~\ref{tab:ew}.
Table shows that they are similar to those for BE381 (Brey 64) that
has $\rm H/He =2 $ and is classified as WN9h by \citet{CrowtherWNh}. 
By analogy with BE381, we suppose that for V532, $\rm H/He \sim 2$. 
It places Romano's star in the region occupied by hydrogen-rich late WN
stars that have atmospheres moderately contaminated by helium.

\begin{table*}\caption{Equivalent widths of helium lines (in H$\beta$
    units) and helium-to-hydrogen abundance ratios for two late
    hydrogen-rich WNs and V532 (2007).}\label{tab:ew}
\center{
\begin{tabular}{lllccccc}
\toprule
               &     &          &  HeI+H8 & H$\gamma$    &  HeII       &    HeI   &  HeI       \\
               &     &          &  3889   &              &  4686       &   4713   &  5876     \\
\midrule
R84            & WN9 & H/He=2.5 &   3    &   2.3        &  3.5        &   12.6   &                      \\
BE381          & WN9h& H/He=2   &   2.1   &   2.4        &  1.9        &   16.6   &  0.8                  \\
V532           &     &          &   2    &    2.75      &  1.4        &   13.4   &  0.8                   \\
\end{tabular}
}
\end{table*}
\bigskip 

We have already mentioned the strikingly different behaviour of
triplet lines of HeI having P~Cyg-like profiles and singlet emissions.
Probably singlet lines have weak absorptions that we can not detect due 
to insufficient signal-to-noise or spectral resolution. 
Besides this, HeI$\lambda$5015 is contaminated by the oxygen [OIII]$\lambda$5007 emission.

Similarity of the line profiles of singlet HeI, [OIII]$\lambda\lambda 4959, 5007$ 
and [ArIII]$\lambda7136$ suggests that both singlet HeI and forbidden
emissions are produced mainly in the low-density ejecta, expanding
at velocities of about 100\kms.
Probably, singlet and triplet lines of neutral helium are formed in different places.    
Because the lowest possible level for triplet lines is metastable,
concentrations of atoms at the lower levels of triplet transitions (at
therefore the intensities of absorption line components) should be
higher. 

We classify the observed evolution of the object as S~Dor
variability cycle. The main difference with AG~Car and other LBVs
is that in the optical minimum V532 becomes much hotter and
may be classified as a WN9/WN8 star, while AG~Car stops at about
WN11. Non-monotonous spectral changes and variability
timescales indicate that these spectral changes hardly correspond to
any real stellar evolution, rather being connected to the ordinary
LBV variability. 
It also means that LBV stars may have spectral
classes as early as WN8. To our notion, there is only one example of
an object that acquired even earlier spectral classes of WN6/7 during
LBV variability cycle, HD5980 \citep{barba,koen08} in
SMC. The hot spectra of this object are probably connected to lower
ambient metallicity, too.

V532 is one more example of an LBV star temporarily becoming a WN.
We predict that, vice versa, some of the Galactic or extra-galactic
Wolf-Rayet stars may prove to be dormant LBVs. Probably, the hottest
possible temperature for an LBV decreases with metallicity, but more
data on the brightest stars in different environments are needed. 

\section{Conclusions}\label{sec:con}
     
Our results show that the object changes from a B emission line
supergiant in the optical maximum, through Ofpe/WN (WN10,WN11)
to WN9 and  further towards a WN8 star in deep minimum.
We confirm the result of \citet{viotti2007}  
 that there is an anti-correlation between the visual luminosity and the temperature of
the star, at larger amount of spectral data. 

V532 spans a large range of spectral classes, becoming one of the
first (probably, the second, after HD5980) LBV stars
noticed to make an excursion as deep as WN8 into the Wolf-Rayet
domain. 
 It is interesting to trace further the evolution of the object. 
 Further monitoring of V532 as well as other late WN stars is
 necessary to establish the evolutionary link between WRs and
 LBVs. Some WNs may prove to be dormant LBVs. 

\acknowledgements
   We are grateful to Vitalij Goranskij, Elena Barsukova and Alla Zharova for providing us with photometric data and Olga
Sholukhova and Thomas Szeifert for the spectrum obtained
with TWIN Calar Alto spectrograph in 1992. We would also like to thank
the anonymous referee for valuable comments and drawing our attention
to HD5980.

\end{document}